\documentclass[reprint,amsmath,amssymb,nofootinbib,
]{revtex4-2}
\usepackage{xtab,afterpage}
\usepackage{graphicx}
\usepackage{dcolumn}
\usepackage{bm}
\usepackage{natbib}
\usepackage{url}
\usepackage{textcase}
\usepackage{amsmath,amssymb,amsfonts,mathrsfs}
\usepackage{mathtools}
\usepackage{braket}
\usepackage{enumitem} 
\usepackage{makecell}
\usepackage{cases}
\usepackage{ulem}
\usepackage{babel}
\usepackage{amsthm}

\newtheorem*{prop}{\uline{Proposition}}
\newtheorem*{theorem}{\uline{Conjecture}}

\newcommand{\f}[2]{\frac{#1}{#2}}
\newcommand{\ko}[1]{\left(#1\right)}
\newcommand{\kko}[1]{\left[#1\right]}

\newcommand{\abs}[1]{\left|#1\right|}

\newcommand{\q}[1]{`#1'}
\newcommand{\dd}{\mathop{}\!d}
\DeclareMathOperator*{\argmin}{arg\,min}

\def\al{\alpha}
\def\be{\beta}

\def\lam{\lambda}
\def\sig{\sigma}
\def\cM{\mathcal{M}}
\def\cH{\mathcal{H}}
\def\cA{\mathcal{A}}
\def\cD{\mathcal{D}}
\def\cK{\mathcal{K}}
\def\cO{\mathcal{O}}
\def\ep{\epsilon}
\def\bX{\boldsymbol{X}}
\def\no{\nonumber}

\DeclareMathOperator{\M}{M}

\begin{document}

\title{Three invariants of strange attractors derived through hypergeometric entropy}

\author{K.~Okamura}
\email{okamura@ifi.u-tokyo.ac.jp}
\affiliation{Institute for Future Initiatives, The University of Tokyo, 7-3-1 Hongo, Bunkyo-ku, Tokyo 113-0033, Japan.}
\date{March 31, 2023}

\begin{abstract}
A new description of strange attractor systems through three geometrical and dynamical invariants is provided.
They are the correlation dimension ($\mathcal{D}$) and the correlation entropy ($\mathcal{K}$), both having attracted attention over the past decades, and a new invariant called the correlation concentration ($\mathcal{A}$) introduced in the present study.
The correlation concentration is defined as the normalised mean distance between the reconstruction vectors, evaluated by the underlying probability measure on the infinite-dimensional embedding space.
These three invariants determine the scaling behaviour of the system's R\'{e}nyi-type extended entropy, modelled by Kummer's confluent hypergeometric function, with respect to the gauge parameter ($\rho$) coupled to the distance between the reconstruction vectors.
The entropy function reproduces the known scaling behaviours of $\mathcal{D}$ and $\mathcal{K}$ in the `microscopic' limit $\rho\to\infty$ while exhibiting a new scaling behaviour of $\mathcal{A}$ in the other, `macroscopic' limit $\rho\to 0$.
The three invariants are estimated simultaneously via nonlinear regression analysis without needing separate estimations for each invariant.
The proposed method is verified through simulations in both discrete and continuous systems.

\end{abstract}

\maketitle

\section{Introduction}

The underlying dynamics of complex phenomena observed in nature are often nonlinear and chaotic, with no analytical solutions available in general.
Usually, the best scientists and engineers can cope with this situation is to identify characteristic statistics from finite time series data of few variables.
The most widely used nonlinearity measure to analyse nonlinear time series data is the fractal dimension introduced by Mandelbrot \cite{Mandelbrot77,Mandelbrot82}.
It offers information about the geometrical (static) property of the system under study, measuring the effective number of degrees of freedom involved in the dynamical process.
There have been proposed many different definitions for the fractal dimension, such as Hausdorff, box-counting (or Minkowski--Bouligand; $D_{0}$), Kaplan--Yorke (KY) (or Lyapunov; $D_{\mathrm{KY}}$) \cite{Kaplan79}, information ($D_{1}$) \cite{Renyi59} and correlation ($D_{2}$) \cite{Grassberger83a,Grassberger83b} dimensions.
Particularly, the correlation dimension has been most widely used in both natural and social sciences for its computational tractability.

Another useful tool to analyse nonlinear time series data is entropy statistics.
It quantifies the unpredictability of the dynamical system, measuring the rate at which nearby trajectories in a phase space diverge as the chaotic system evolves in time.
A sufficient condition for chaos is that the entropy is positive.
The definition of the entropy is not unique either, and several different entropy measures are commonly used in literature, including Kolmogorov--Sinai (KS) (or metric; $K_{1}$) \cite{Kolmogorov58,Sinai59} and correlation ($K_{2}$) \cite{Termonia84,Szepfalusy86,Redaelli02,Urbanowicz03} entropies.
Due to some technical difficulty and computational cost \cite{Harikrishnan09}, these entropy measures have been used less extensively in practical applications compared to the fractal dimensions. 
Still, their utility and importance as discriminative statistics to characterise nonlinear dynamical systems are no less significant than the fractal dimension, especially in the context of coloured noise contamination \cite{Szepfalusy86,Redaelli02,Urbanowicz03}.

In addition, Lyapunov exponents \cite{Lyapunov92} provide an essential tool for quantitative analysis of the dynamics of chaotic states.
They are defined as the logarithms of the eigenvalues of the product of all Jacobian matrices along an infinitely long trajectory.
There have been established several remarkable conjectures on the relations between the chaotic invariants and the Lyapunov spectrum $\{\lam_{\ell}\}_{\ell=1,\,2,\,\dots}$. 
In particular, KY conjecture \cite{Kaplan79,Frederickson83} states that the fractal dimension (specifically, $D_{1}$) is equal to the KY dimension defined by $D_{\mathrm{KY}}=g+\sum_{\ell=1}^{g}\lam_{\ell}\big/\abs{\lam_{g+1}}$.
Here, $g$ is the largest integer for which $\sum_{\ell=1}^{g}\lam_{\ell}\geq 0$, where the exponents are arranged in decreasing order (i.e., $\lam_{\ell}\geq\lam_{\ell'}$ for $\ell<\ell'$).
It has been proven \cite{Ledrappier81} that $D_{\mathrm{KY}}$ provides an upper bound on $D_{1}$.
Also, Pesin's formula \cite{Pesin77} makes a connection between the system's entropy and the Lyapunov exponents.
It states that for a closed ergodic system, the KS entropy is equal to the sum of the positive Lyapunov exponents, integrated with respect to the associated invariant probability measure $\mu$; i.e., $K_{1}(\mu)=\int\dd\mu\sum_{\ell:\,\lam_{\ell}>0}m_{\ell}\lam_{\ell}$, where $m_{\ell}$ is the multiplicity of the $\ell$-th exponent.

This paper further explores the relation between the geometrical and dynamical invariants of chaotic systems and the underlying probability measure.
We mainly focus on strange chaotic attractor systems.
They are \q{strange} because they are fractal in nature, exhibiting locally unstable yet globally stable with self-similar and scale-invariant behaviour. 
A new description of strange attractor systems in terms of three chaotic invariants is provided through a R\'{e}nyi-type extended entropy function.
The extended entropy is derived from a modified correlation integral, in which the kernel function is given by a smooth exponential function, rather than the Heaviside function used in the Grassberger--Procaccia (GP) method \cite{Grassberger83a,Grassberger83b}.

In Section \ref{sec:theory}, we introduce an estimator of the modified correlation integral and show how it can be used as a unique probe for the spatiotemporal properties of strange attractors.
The new estimator takes the form of Kummer's confluent hypergeometric function, whose parameters involve the three chaotic invariants.
In Section \ref{sec:simulation}, the proposed method is verified through simulations in six strange attractors (logistic and H\'{e}non maps; Lorenz, R\"{o}ssler, Duffing--Ueda and Langford attractors).
The estimates of the three chaotic invariants are obtained simultaneously through nonlinear regression.
Section \ref{sec:discussion} is devoted to discussions on how the structure of the underlying probability measure can be described through the chaotic invariants.
Finally, the summary and conclusions are presented in Section \ref{sec:summary}.

\section{Extended theory and the correlation concentration\label{sec:theory}}

\subsection{Scaling law at microscopic scale and Grassberger--Procaccia algorithm}

As a general illustration, let us consider an attractor $\mathscr{A}$ whose fractal structure is evolving in a phase space with embedding dimension $m\in \mathbb{Z}_{+}$.
It can be either discrete ($t\in \mathbb{Z}_{+}$) or continuous ($t\in \mathbb{R}_{+}$), depending on how the dynamical trajectory $\{(x(t),y(t),z(t),\dots)\}$ in the phase space is described as a vector function of time ($t$).
The Taken's embedding theorem \cite{Takens81} states that if the embedding dimension is at least twice the box-counting dimension of the attractor ($m>2D_{0}$), then the reconstructed attractor is diffeomorphic to the original attractor \cite{Sauer91}.
The entire attractor can be reconstructed from a one-dimensional time series $\{x(t_{i})\}_{i=0}^{n-1}$, $t_{i}=t_{0}+i\tau$, where $t_{0}$ and $\tau$ represent the initial time and the sampling time interval, respectively.
Subsequently, $m$-dimensional reconstruction vectors are defined via the method of time-delay coordinates \cite{Takens81} as
\begin{equation}\label{reconvec}
\bX_{i}\coloneqq \big[x(t_{i}),\,x(t_{i+1}),\,\dots,\,x(t_{i+m-1})\big]^{\intercal}\in\mathbb{R}^{m}\,.
\end{equation}

The GP method \cite{Grassberger83a,Grassberger83b,Hentschel83,Pawelzik87} provides a standard algorithm to obtain the estimates of the correlation dimension and entropy from the reconstructed phase space.
A probabilistically-defined quantity called the correlation integral (or the correlation sum, at finite $n$) plays a central role in this approach.
At partition scale $\ep$, it is conceptualised as the probability that two points randomly drawn from $\mathscr{A}$ according to the $m$-dimensional probability measure $\mu_{m}$ are within a mutual distance of $\ep$.
It is formally defined by 
\begin{align}\label{C_m}
{C}_{m}(\ep)&\coloneqq\iint_{\mathscr{A}\times\mathscr{A}}\dd\mu_{m}(\bX_{i},\bX_{j})\,\Theta(\ep-\|\bX_{i}-\bX_{j}\|)\no\\
&=\lim_{n\to\infty}\f{\sum_{i\neq j}\Theta(\ep-\|\bX_{i}-\bX_{j}\|)}{n(n-1)}\,,
\end{align}
where $\|\cdot\|$ is the $L^{p}$-norm defined by $\|\bX_{i}-\bX_{j}\|\coloneqq \big(\sum_{\ell=0}^{m-1}\abs{x(t_{i+\ell})-x(t_{j+\ell})}^{p}\big)^{1/p}$ with $p\geq 1$, and $\Theta$ is the standard Heaviside function defined such that $\Theta(x)=1$ if $x\geq 0$ and $\Theta(x)=0$ if $x<0$.
Although the correlation dimension and entropy are invariant with respect to the choice of the norm, hereafter we use the maximum norm ($p=\infty$) for later purpose, i.e.,
\begin{equation}
\|\bX_{i}-\bX_{j}\|\coloneqq \max_{\ell=0,\,\dots,\,m-1}\{\abs{x(t_{i+\ell})-x(t_{j+\ell})}\}\,.
\end{equation}
In evaluating the correlation sum, we restrict our analysis to its \q{off-diagonal} components than directly analysing the entire correlation sum, including the diagonal ($i=j$) components.
Namely, we employ the $U$-statistic rather than the $V$-statistic.
In addition, the probability measure is replaced by the so-called empirical reconstruction measure \cite{Diks99}, which assigns equal mass, $1/n$, to each of the reconstruction vectors.

Subsequently, the information on the correlation dimension ($D_{2}$) and the correlation entropy ($K_{2}$) can be extracted by taking the simultaneous limit $\ep\to 0$ and $m\to \infty$, in which the correlation integral (\ref{C_m}) behaves as 
\begin{equation}\label{C:scaling}
C_{m}(\ep)\, \sim\, \ep^{D_{2}}e^{-K_{2}m\tau}\,.
\end{equation}
The scaling property with respect to the space variable ($\ep$) characterises the fractal structure of the support of $\mu_{m}$,
while the rate of change with respect to the time variable ($m\tau$) characterises the dynamical property of the system.
The correlation dimension and entropy can then be formally defined by
\begin{align}
D_{2}&\coloneqq\lim_{\ep\to 0}\lim_{m\to \infty}\f{\ln C_{m}(\ep)}{\ln\ep}\,,\\
K_{2}&\coloneqq\lim_{\ep\to 0}\lim_{m\to \infty}\f{1}{\tau}\ln\ko{\f{C_{m}(\ep)}{C_{m+1}(\ep)}}\,.
\end{align}
Note that neither of the limits, $\ep\to 0$ or $m\to \infty$, can be achieved in practice since experimental data is always of finite length and finite resolution.

To estimate $D_{2}$ and $K_{2}$ from the experimental time series data, the GP method first identifies a linear part $\mathscr{R}=[\ep_{1},\ep_{2}]$ in the $\ln C_{m}$ versus $\ln\ep$ plot to avoid outliers and edge effects.
The scaling region is identified by visual inspection, and therefore, it is necessarily subjective.
The slope of the plot in $\mathscr{R}$ is obtained by the least square fitting through the points $\{(\ln\ep_{i},\ln C_{m}(\ep_{i}))\}$ for each $m$.
The correlation dimension is estimated by taking the mean of the slopes over \q{well-behaved} embedding dimensions, 
for which the estimated slopes remain relatively constant.
Also, the correlation entropy is estimated by the relation $K_{2}(\ep_{i})=[\ln C_{m}(\ep_{i})-\ln C_{m+1}(\ep_{i})]\big/\tau$ in the same scaling region $\mathscr{R}$, by first taking the mean over $i$, and then over $m$.

The above GP method and the variations thereof \cite{Grassberger83a,Grassberger83b,Hentschel83,Pawelzik87,Grassberger88} thus look into the \q{microscopic} structure of the system in the region $\ep\ll 1$ to obtain the estimates of $D_{2}$ and $K_{2}$.
Then, what if we look at the other, \q{macroscopic} limit, in which $\ep$ is sufficiently large?
Of course, the correlation integral saturates to $C=1$ if $\ep\geq \max_{i,j}\{\abs{x(t_{i})-x(t_{j})}\}$.
As the value of $\ep$ decreases from the maximum, the curve profile of $\ln C_{m}$ versus $\ln\ep$ behaves nontrivially, reflecting the nontrivial distribution of the reconstruction vectors.
In the following section, we derive a new characteristic statistic ($\cA^{*}$) from this \q{macroscopic} region through a modification of the correlation integral.
Indeed, the correlation integral does not generally allow an analytical expression because of the nontrivial structure of the associated probability measure.
However, we show that our modified correlation integral does allow an analytical expression by a simple nonlinear function, specifically Kummer's confluent hypergeometric function.
It reproduces the known scaling behaviours of ${D}_{2}$ and ${K}_{2}$ in the small-$\ep$ region while exhibiting a new scaling behaviour of $\cA^{*}$ in the large-$\ep$ region.
Notably, these three invariants can be estimated simultaneously via nonlinear regression analysis without needing separate estimations for each invariant.

\subsection{New scaling law at macroscopic scale via extended correlation integral\label{sec:cc}}

Let us modify the correlation integral of Eq.~(\ref{C_m}) by replacing the Heaviside function with an exponential function of the form $e^{-\rho\|\bX_{i}-\bX_{j}\|}$, with $\rho$ a positive gauge parameter.
In the $\rho\to\infty$ limit, this parameter effectively plays the role of $\ep^{-1}$ of the GP method.
This replacement of the kernel function leads to the following extended correlation integral, 
\begin{align}\label{M_m}
\cM_{m}(\rho)&\coloneqq\iint_{\mathscr{A}\times\mathscr{A}}\dd\mu_{m}(\bX_{i},\bX_{j})\,e^{-\rho\|\bX_{i}-\bX_{j}\|}\no\\
&=\lim_{n\to\infty}\f{\sum_{i\neq j}e^{-\rho \|\bX_{i}-\bX_{j}\|}}{n(n-1)}\,.
\end{align}
This function of $\rho$ can be seen as the normalised moment-generating function of the distance between the reconstruction vectors.
This extension of the correlation integral may appear engineeringly costly but has an essential physical and information-theoretical meaning.
Specifically, the extended correlation integral (\ref{M_m}) is related to the so-called entropy power (or the Uffink's class of entropies, or the diversity index) \cite{Shannon48,Hill73,Uffink95,Jizba19}.
For a discrete probability variable $p_{i}\in [0,1]$, satisfying $\sum_{i}p_{i}=1$, it is given by ${}^{q}\!\mathit{\Delta}(\{p_{i}\})=\big(\sum_{i}(p_{i})^{q}\big)^{\f{1}{1-q}}$ for $q\neq 1$ and ${}^{1}\!\mathit{\Delta}(\{p_{i}\})=e^{-\sum_{i}p_{i}\ln p_{i}}$ for $q=1$, where $q$ is called the order parameter.
It holds that $\lim_{q\to 1}{}^{q}\!\mathit{\Delta}={}^{1}\!\mathit{\Delta}$.
Recently, it has been generalised to the following distance-based extended form \cite{Okamura20},
\begin{align}\label{KO:discrete}
\resizebox{.48\textwidth}{!} 
{$
{}^{q}\!\mathit{\Delta}(\{p_{i}\},\{d_{ij}\})
=\begin{cases}
\,\Big[\sum_{i}p_{i}\big(\sum_{j}p_{j}\kappa(d_{ij})\big)^{q-1}\Big]^{\f{1}{1-q}} & (q\neq 1) \\[2mm]
\,\exp\big[-\sum_{i}p_{i}\ln\sum_{j}p_{j}\kappa(d_{ij})\big]  & (q=1)
\end{cases}\,.
$}
\end{align}
Here, $d_{ij}\in [0,1]$ denotes a certain properly-defined distance between the $i$-th and the $j$-th elements.
The kernel function $\kappa(\cdot)$ is a monotonically decreasing function satisfying $\kappa(0)=1$ and $\kappa(1)=0$.
The generalisation to the case of a continuous system is straightforward, the correspondence being $p_{i}\leftrightarrow\mu(d\bX_{i})$, $d_{ij}\leftrightarrow\rho\|\bX_{i}-\bX_{j}\|$ and $\kappa(d_{ij})\leftrightarrow e^{-\rho\|\bX_{i}-\bX_{j}\|}$.
Subsequently, the entropy power associated with an attractor $\mathscr{A}$ is given by
\begin{align}\label{KO:continuous}
\resizebox{.48\textwidth}{!} 
{$
{}^{q}\!\mathit{\Delta}_{m}(\mu;\rho)
=\Bigg[ \int_{\mathscr{A}}\dd\mu_{m}(\bX_{i})
\bigg( \int_{\mathscr{A}}\dd\mu_{m}(\bX_{j})\,
e^{-\rho\|\bX_{i}-\bX_{j}\|}\bigg)^{q-1} \Bigg]^{\f{1}{1-q}}
$}
\end{align}
for $q\neq 1$, and ${}^{1}\!\mathit{\Delta}_{m}(\mu;\rho)\coloneqq\lim_{q\to 1}{}^{q}\!\mathit{\Delta}_{m}(\mu;\rho)$ for $q=1$.
Note that different values of $q$ yield different entropy measures, enabling to probe the multifractality of the system.
Specifically, the $q=1$ case corresponds the distance-based generalised KS entropy, and the $q=2$ case corresponds to the current extended correlation integral (\ref{M_m}); i.e., ${}^{2}\!\mathit{\Delta}_{m}(\mu;\rho)=\cM_{m}^{-1}(\rho)$.

Observing that the small-$\ep$ region of the GP correlation integral (\ref{C_m}) corresponds to the large-$\rho$ region of Eq.~(\ref{M_m}), the scaling law (\ref{C:scaling}) now translates to
\begin{equation}\label{scaling:DK}
\cM_{m}(\rho)\, \sim\, 
\rho^{-\cD^{*}}e^{-\cK^{*}m\tau}\quad
\text{as}~\, \rho\to \infty,\,m\to \infty\,.
\end{equation}
Here, $\cD^{*}$ and $\cK^{*}$ denote the correlation dimension and entropy, respectively.
We have used calligraphic letters to indicate that they are derived from the extended correlation integral of Eq.~(\ref{M_m}), rather than the original Eq.~(\ref{C_m}).
The extended correlation integral naturally introduces a new natural measure that cannot be captured by the GP method.
Consider the small-$\rho$ expansion of the extended correlation integral,
\begin{equation}\label{scaling:A}
\cM_{m}(\rho)\, \sim\, 
1-\cA_{m}\rho\quad
\text{as}~\, \rho\to 0,\,m\to \infty\,.
\end{equation}
Here, $\cA_{m}$ represents the mean distance between the reconstruction vectors $\{\bX_{i}\}$, evaluated by the probability measure $\mu_{m}$ on the $m$-dimensional embedding space, i.e.,
\begin{align}
\cA_{m}&\coloneqq\iint_{\mathscr{A}\times\mathscr{A}}\dd\mu_{m}(\bX_{i},\bX_{j})\,\|\bX_{i}-\bX_{j}\|\no\\
&=\lim_{n\to\infty}\f{\sum_{i\neq j}\|\bX_{i}-\bX_{j}\|}{n(n-1)}\,,
\end{align}
which clearly depends on the norm $\|\cdot\|$ and the embedding dimension $m$.
As $m$ increases, $\cA_{m}$ is expected to approach asymptotically to a constant value.
Based on this observation, we define what we call the \textit{correlation concentration} by
\begin{align}\label{def:A}
\cA^{*}\coloneqq\lim_{m\to \infty}\f{\cA_{m}}{\max_{i,j}\{\|\bX_{i}-\bX_{j}\|\}}\,.
\end{align}
It is a simple quantity, yet it proves to be a useful characteristic in analysing nonlinear time series data.
As discussed in Appendix \ref{app:cc}, for a uniform random variable case (Fig.~\ref{fig:topolo}(a)), the correlation concentration is shown to be given by $\cA^{*}_{\mathrm{uni}}=1$, giving the upper bound of the correlation concentration for any system.
By contrast, for a deterministic (straight-line) case (Fig.~\ref{fig:topolo}(b)), it is shown to be given by $\cA^{*}_{\mathrm{line}}=1/3$.
Now, strange attractors are neither similar to completely random systems nor perfectly predictable systems (like straight lines or periodic orbits) in terms of the scaling behaviour of the correlation integral.
The difference arises from the nontrivial, intrinsic geometrical and dynamical structure of the underlying probability measure; see Fig.~\ref{fig:topolo}(c) for an illustration of H\'{e}non map.
Here, the correlation concentration $\cA^{*}$, in addition to $\cD^{*}$ and $\cK^{*}$, can provide a useful discriminative statistic to characterise the system.
Already at the intuitive level, the value of $\cA^{*}$ for a strange attractor is expected to be lower than $\cA^{*}_{\mathrm{uni}}~(=1)$ but higher than $\cA^{*}_{\mathrm{line}}~(=1/3)$.
Below we explain how our method, based on the extended correlation integral (\ref{M_m}), provides information of $\cA^{*}$ and its interrelation with $\cD^{*}$ and $\cK^{*}$.

Before doing so, we comment on previous studies that also considered extensions of the GP estimator by replacing the \q{hard} Heaviside function with some \q{soft} kernel functions \cite{Diks96,Yu00,Nolte01}.
For instance, Diks et al.\ \cite{Diks96} chose the kernel function to be of the Gaussian form, $e^{-\|\bX_{i}-\bX_{j}\|^{2}/4\ep^{2}}$ with $\|\cdot\|$ the $L^{2}$-norm.
This Gaussian kernel approach is useful for capturing the effect of additive Gaussian noise on estimated correlation dimensions.
Specifically, for a time series with added white Gaussian noise of variance $\tilde{\sig}^{2}$, the scaling law is modified to $C_{m}(\ep,\tilde{\sig})\sim \ep^{m}\big(\ep^{2}+\tilde{\sig}^{2}\big)^{\f{D_{2}-m}{2}}e^{-K_{2}m\tau}$ in the simultaneous limit $m\to\infty$ and $\ep^{2}+\tilde{\sig}^{2}\to 0$ \cite{Diks96,Yu00,Nolte01}.
By contrast, our use of the kernel function of the form $e^{-\rho\|\bX_{i}-\bX_{j}\|}$ is aimed at obtaining an improved understanding of the geometrical and dynamical properties of noise-free strange attractors, obeying the pure scaling law (\ref{scaling:DK}).

\subsection{Interpolating between micro- and macro-scales of chaos through hypergeometric lens}

Based on the extended correlation integral (\ref{M_m}), we define the R\'{e}nyi-type extended entropy by
\begin{equation}\label{def:H}
\cH_{m}(\rho)\coloneqq -\ln\cM_{m}(\rho)\,.
\end{equation}
We also introduce the logarithmic scale parameter defined by $\sig\coloneqq\ln\rho$.
Then, the scaling laws (\ref{scaling:DK}) and (\ref{scaling:A}) translate to, respectively,
\begin{numcases}{\cH_{m}(\sig) \sim }
      \, \cD^{*}\sig+\cK^{*}m\tau &$\text{as}~\, \sig\to \infty,~m\to\infty$\,,\label{scaling:sig_DK}\\
     \, \cA_{m}e^{\sig} &$\text{as}~\, \sig\to -\infty,~m\to \infty$\,,\label{scaling:sig_A}
\end{numcases}
where $\cA_{m}$ and $\cA^{*}$ are related by Eq.~(\ref{def:A}).
Subsequently, the correlation dimension, entropy and concentration are formally defined as
\begin{align}
\cD^{*}&=\lim_{\sigma\to \infty}\lim_{m\to \infty}\partial_{\sigma}\cH_{m}(\sigma)\,,\label{D*0}\\
\cK^{*}&=\lim_{\sigma\to \infty}\lim_{m\to \infty}\kko{\cH_{m+1}-\cH_{m}}(\sigma)\big/\tau\,,\label{K*0}\\
\cA^{*}&=\lim_{\sig\to -\infty}\lim_{m\to \infty}e^{-\sig}\partial_{\sig}\cH_{m}(\sig)\,.
\label{A*0}
\end{align}
It is straightforward to generalise the above results to the general $q$ case as follows:
\begin{prop}
Let $\mathscr{A}$ be an attractor of a dynamical system, $m\in \mathbb{Z}_{+}$ be the embedding dimension, and $\mu_{m}$ be the $m$-dimensional probability measure.
Let also $\rho\in\mathbb{R}_{+}$ be the gauge parameter, and $q\in\mathbb{R}$ be the order parameter.
Then the correlation dimension, entropy and concentration of order-$q$ associated with $\mathscr{A}$ are given by, respectively,
\begin{align}
\cD^{*}_{q}&=\f{1}{q-1}\lim_{\sigma\to \infty}\lim_{m\to \infty}\partial_{\sigma}\cH_{m,q}(\sigma)\,,\label{D*0,q}\\
\cK^{*}_{q}&=\f{1}{\tau(q-1)}\lim_{\sigma\to \infty}\lim_{m\to \infty}\kko{\cH_{m+1,q}-\cH_{m,q}}(\sigma)\,,\label{K*0,q}\\
\cA^{*}_{q}&=(q-1)\lim_{\sig\to -\infty}\lim_{m\to \infty}e^{-\sig}\partial_{\sig}\cH_{m,q}(\sig)\,,
\label{A*0,q}
\end{align}
where
\begin{align}\label{def-q}
\resizebox{.48\textwidth}{!} 
{$
\cH_{m,q}(\rho)
=\f{1}{1-q}\ln\int_{\mathscr{A}}\dd\mu_{m}(\bX_{i})
\Big( \int_{\mathscr{A}}\dd\mu_{m}(\bX_{j})\,
e^{-\rho\|\bX_{i}-\bX_{j}\|} \Big)^{q-1}
$}
\end{align}
with $\{\bX_{i}\}_{i=0}^{n-1}$ the set of $m$-dimensional reconstruction vectors defined by Eq.~(\ref{reconvec}).
The case of $q=1$ is understood as the limit as $q\to 1$.
\end{prop}

\begin{figure}[t]
 \centering
\includegraphics[width=0.48\textwidth]{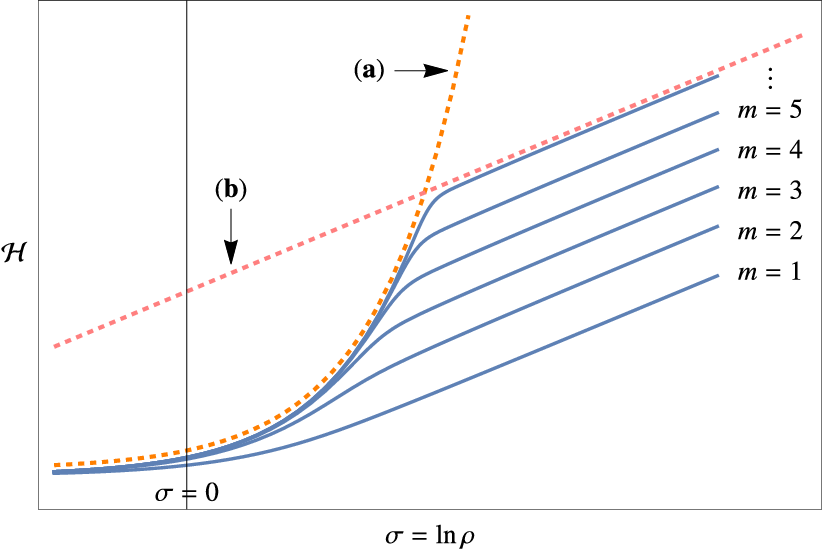}
\caption{%
Schematic plot of the extended R\'{e}nyi entropy $\cH_{m}$ versus $\sig=\ln\rho$.
The curve consists of (a) an exponential part in the $\rho\ll 1$ region (Eq.~(\ref{scaling:sig_A})) and (b) a linear part in the $\rho\gg 1$ region (Eq.~(\ref{scaling:sig_DK})).
The correlation concentration, together with the correlation entropy, determines the curvature of the exponential part.
The correlation dimension determines the slope of the linear part of the curve, whereas the correlation entropy determines the vertical gap between the consecutive linear segments.
\label{fig:wing}}
\end{figure}

Returning to the case of $q=2$, an interesting question to ask here is what the exact functional form of $\cH_{m}(\sigma)$ satisfying the scaling laws (\ref{scaling:sig_DK}) and (\ref{scaling:sig_A}) is.
If such an analytical expression of the entropy function is available, it is not only theoretically intriguing but also practically effective.
If one could conduct a curve-fitting for the whole domain of $\rho$ and simultaneously obtain the estimates of these chaotic invariants, it is statistically more preferable than conducting a separate curve-fitting for $\rho\gg 1$ (to obtain $\cD^{*}$ and $\cK^{*}$) and $\rho\ll 1$ (to obtain $\cA^{*}$ and $\cK^{*}$) for each $m$.
Indeed, it is impossible to write down the exact analytical form of $\cH_{m}(\sigma)$ without knowledge of the underlying probability measure $\mu_{m}$.
However, it is still possible to derive a nonlinear function of the spatiotemporal parameters $\sig$ and $m\tau$ that satisfies the limiting behaviours (\ref{scaling:sig_DK}) and (\ref{scaling:sig_A}) in the respective limits ($\sig\to\pm\infty$) of the scale parameter at large $m$.
The key to this approach is the use of Kummer's confluent hypergeometric function, defined by
\begin{equation*}
\M(a,b;x)\coloneqq\sum_{k=0}^{\infty}\f{(a)_{k}}{(b)_{k}}\f{x^{k}}{k!}\,,
\end{equation*}
where $a,\,b\,,x\in \mathbb{R}$, and $(a)_{k}$ denotes the Pochhammer polynomial (or the rising factorial) given by $(a)_{k}=a(a+1)\cdots (a+k-1)=\Gamma(a+k)/\Gamma(a)$ for $k\in\mathbb{Z}_{+}$.
This function, also often denoted as ${}_{1}F_{1}(a,b;x)$, commonly appears in a wide range of classical and quantum physics, chemistry, and engineering disciplines \cite{Mathews22}.
We claim that the extended correlation integral, Eq.~(\ref{M_m}), can be modelled and approximated by 
\begin{align}\label{M:est}
\widetilde{\cM}_{m}(\sig)=\M[\cD^{*},\cD^{*}(1+e^{-\cK^{*} m\tau});-\cA^{*} e^{\sig}]\,.
\end{align}
The corresponding R\'{e}nyi entropy $\widetilde{\cH}_{m}(\sig)$ is defined by the relation (\ref{def:H}).
Here and hereafter, the tilde denotes the approximated function.
We can check that this function precisely obeys the scaling laws of Eqs.~(\ref{scaling:sig_DK}) and (\ref{scaling:sig_A}).
Specifically, it can be shown that, for the \q{microscopic} limit,
\begin{equation}\label{lim:H_DK}
\widetilde{\cH}_{m}(\sig)\,\sim\,\cD^{*}\sigma+\cK^{*}m\tau +\text{const.}
\quad\text{as}~\, \sig\to \infty,\, m\to \infty\,,
\end{equation}
reproducing the scaling behaviour (\ref{scaling:sig_DK}).
See Appendix \ref{app:derivation} for details of the derivation.
In addition, it is easy to see that, for the \q{macroscopic} limit,
\begin{equation}\label{lim:H_A}
\widetilde{\cH}_{m}(\sig)\,\sim\,\f{\cA^{*}e^{\sig}}{1+e^{-\cK^{*}m\tau}}
\quad\text{as}~\, \sig\to -\infty,\, m\to \infty\,,
\end{equation}
which follows from the expansion for small $x$ of the Kummer's function, $\M(a,b;x)\sim 1+(a/b)x+\cO(x^{2})$.
By comparing Eq.~(\ref{lim:H_A}) with Eq.~(\ref{scaling:sig_A}), the probabilistic mean distance between the reconstruction vectors embedded in $\mathbb{R}^{m}$ is identified to be
\begin{equation}\label{A-Am}
\cA_{m}\equiv\f{\cA^{*}}{1+e^{-\cK^{*}m\tau}}\,.
\end{equation}

Figure \ref{fig:wing} shows a schematic plot of the extended entropy $\widetilde\cH_{m}$ versus $\sig=\ln\rho$ for increasing values of the embedding dimension ($m=1,\,2,\,\dots$).
The curve profile resembles a bird's wing in shape; it consists of an exponential, \q{concave up}-shape part ($\rho\ll 1$) plus a positively-sloped linear part ($\rho\gg 1$).
The correlation dimension ($\cD^{*}$) is represented by the slope of the linear part, whereas the correlation entropy ($\cK^{*}$) (times the sampling time interval) is represented by the vertical gap between the consecutive linear segments.
In addition, the correlation concentration ($\cA^{*}$) and entropy ($\cK^{*}$) determine the curvature of the exponential part.

There are several advantages of using the above hypergeometric estimator over the GP linear estimator.
First, its interpolating property regarding the gauge parameter $\rho$ enables us to investigate the spatiotemporal properties of the chaotic system at both the \q{microscopic} ($\rho\to\infty$) and the \q{macroscopic} ($\rho\to 0$) scales.
Consequently, not only $\cD^{*}$ and $\cK^{*}$, which can be obtained by the GP method and characterise the local structure of the chaotic system, but also $\cA^{*}$, characterising the global structure, can be obtained, allowing for a more detailed description of the system.
Second, the estimates of the three invariants are obtained simultaneously via nonlinear regression analysis at a single $m~(\gg 1)$.
This is in contrast to the GP case, where the estimates are necessarily obtained separately for each invariant, $\cD^{*}$ and $\cK^{*}$, where multiple curve-fittings at different values of $m$ are required to obtain the estimate of $\cK^{*}$.
Third, the proposed method can mitigate the subjectivity of setting the scaling region used to compute the correlation dimension and entropy, albeit not entirely but partially.
Recall that the GP method identifies the scaling region $\mathscr{R}=[\ep_{1},\ep_{2}]$ by visual inspection to be where $\ln C(\ep)$ scales linearly with respect to both $\ln\ep$ and $m\tau$.
This method introduces a subjectivity issue for both endpoints, $\ep_{1}$ and $\ep_{2}$.
By contrast, our method does not require the specification of one of the endpoints.
Specifically, for the linear scaling region $[\sig_{1},\sig_{2}]$ in our case, the value of $\sig_{1}$ no longer needs to be specified by hand but is fixed by the requirement of the overall goodness-of-fit of the hypergeometric regression.
Specifically, the value of $\sig_{1}$ is approximately given by the positive solution of the following equation:
\begin{equation*}
\f{\hat{\cA}^{*}e^{\sig_{1}}}{1+e^{-\hat{\cK}^{*}m\tau}}=\hat{\cD}^{*}\sig_{1}+\ln\kko{\f{\hat{\cA}^{*}{}^{\hat{\cD}^{*}}\Gamma\big(\hat{\cD}^{*}e^{-\hat{\cK}^{*}m\tau}\big)}{\Gamma\big(\hat{\cD}^{*}\big(1+e^{-\hat{\cK}^{*}m\tau}\big)\big)}}\,.
\end{equation*}
Here and hereafter, the hat denotes the estimated quantity.

Before demonstrating our approach to extracting the chaotic invariants from a dynamic system, we summarise our main results above as a conjecture.
Again, it is straightforward to generalise the discussion to the general $q$ case as follows:
\begin{theorem}
The order-$q$ entropy function of Eq.~(\ref{def-q}) is analytically given in terms of Kummer’s confluent hypergeometric function by
\begin{align}\label{M:est-q}
\resizebox{.48\textwidth}{!} 
{$
\widetilde{\cH}_{m,q}(\rho)=-\ln\M\bigg[(q-1)\cD^{*}_{q},(q-1)\cD^{*}_{q}(1+e^{-(q-1)\cK^{*}_{q} m\tau});-\f{\cA^{*}_{q} \rho}{q-1}\bigg]\,,
$}
\end{align}
where $\cD^{*}_{q}$, $\cK^{*}_{q}$ and $\cA^{*}_{q}$ are the correlation dimension, entropy and concentration of order-$q$, respectively.
The case of $q=1$ is understood as the limit as $q\to 1$.
\end{theorem}

\section{Hypergeometric curve fitting on chaotic time series \label{sec:simulation}}

\subsection{Method}

We now demonstrate how the proposed method works through concrete examples and verifies its practical usefulness.\footnote{All the simulation results and graphs presented in this paper were produced by Mathematica (Wolfram Research, Inc.).
Its built-in function \texttt{NonlinearModelFit} was used for the nonlinear curve fitting.}
We applied the method to six strange attractors; specifically, two discrete (logistic and H\'{e}non maps) and four continuous (Lorenz, R\"{o}ssler, Duffing--Ueda and Langford attractors) systems.
For all the systems investigated, the left endpoint of the curve-fitting region was set as $\sigma_{0}=-2$, effectively ensuring $\rho\approx 0$.
For the continuous systems, we defined the discretised $x$-variable as $x_{k}\coloneqq x(t_{k})=x(t_{0}+k\tau)$ with $\tau$ the sampling time interval, set as $\tau=1$, and the same for $y_{k}$ and $z_{k}$ (if applies).
The size (total diameter) of the reconstructed attractor was rescaled as $x_{k}\mapsto {x}'_{k}$ such that $\max_{i,j}\{|{x}'_{i}-{x}'_{j}|\}=1$, hence $\cA^{*}=\lim_{m\to\infty}\cA_{m}$.

Subsequently, the phase spaces of the strange attractors were reconstructed by the method of time-delayed coordinates \cite{Takens81}.
By plugging the reconstruction vectors in the formulae (\ref{M_m}) and (\ref{def:H}), we created the plot of the extended R\'{e}nyi entropy $\cH_{m}$ against $\sig=\ln\rho$, discretised by the step of $0.1$.
As discussed, we did not need to specify the value of the internal endpoint $\sig_{1}$, at which the linear part starts.
It is because the fitting curve is smoothly connected to the new scaling behaviour in the small-$\rho$ region, which is controlled by $\cA^{*}$ and $\cK^{*}$.
However, we still needed to specify the value of the external endpoint $\sig_{2}$ for the curve-fitting.
It is because for large values of $\sig\gtrsim\sig_{2}$, deviation from the scaling law becomes visible as statistical fluctuations dominate the estimation; it is a situation where only a few distances contribute to $\cH_{m}$.
For the discrete systems, the value of $\sig_{2}$ was determined independently for each system by visual inspection, while for the continuous systems, it was commonly set as $\sigma_{2}=4.5$.

By fitting the hypergeometric estimator $\widetilde\cH_{m}(\sig)$ derived from Eq.~(\ref{M:est}) on the time series data defined on $[\sig_{0},\sig_{2}]$, the estimates of the three chaotic invariants, $\hat{\cD}(m)$, $\hat{\cK}(m)$ and $\hat{\cA}(m)$, were simultaneously obtained for each $m$.
On increasing the value of $m$, these estimates tend to converge to their respective asymptotic values, $\hat{\cD}^{*}$, $\hat{\cK}^{*}$ and $\hat{\cA}^{*}$.
Theoretically and formally, it holds that $\lim_{m\to\infty}\hat{\cD}(m)=\hat{\cD}^{*}$ and the same for $\hat{\cK}^{*}$ and $\hat{\cA}^{*}$.
However, in the current experimental setting, the estimated values diverge for some $m$ due to decorrelation.
Therefore, the values of $\hat{\cD}^{*}$, $\hat{\cK}^{*}$ and $\hat{\cA}^{*}$ were obtained by averaging over the embedding dimension range where the estimates were relatively independent of $m$.
For the discrete systems, the range of $m$ was determined independently for each system by visual inspection, while for the continuous systems, it was commonly set as $m$=$10$--$20$.

For the discrete systems, we generated a time series consisting of $n=10,000$ consecutive observations of the $x$-variable.
For the continuous systems, the length of the sampling time interval ($n$) was searched in the range $[200,400]$.
It was determined independently for each system such that the resulting estimates of the chaotic invariants, especially that of the correlation dimension $\hat{\cD}^{*}(m)$, remain relatively constant throughout the averaging range of $m$.
If multiple values of $n$ satisfy these criteria, the one with the highest $\hat{\cD}^{*}$ was selected.

\subsection{Results}

\onecolumngrid

\begin{figure}[t]
\centering
\includegraphics[width=0.85\textwidth]{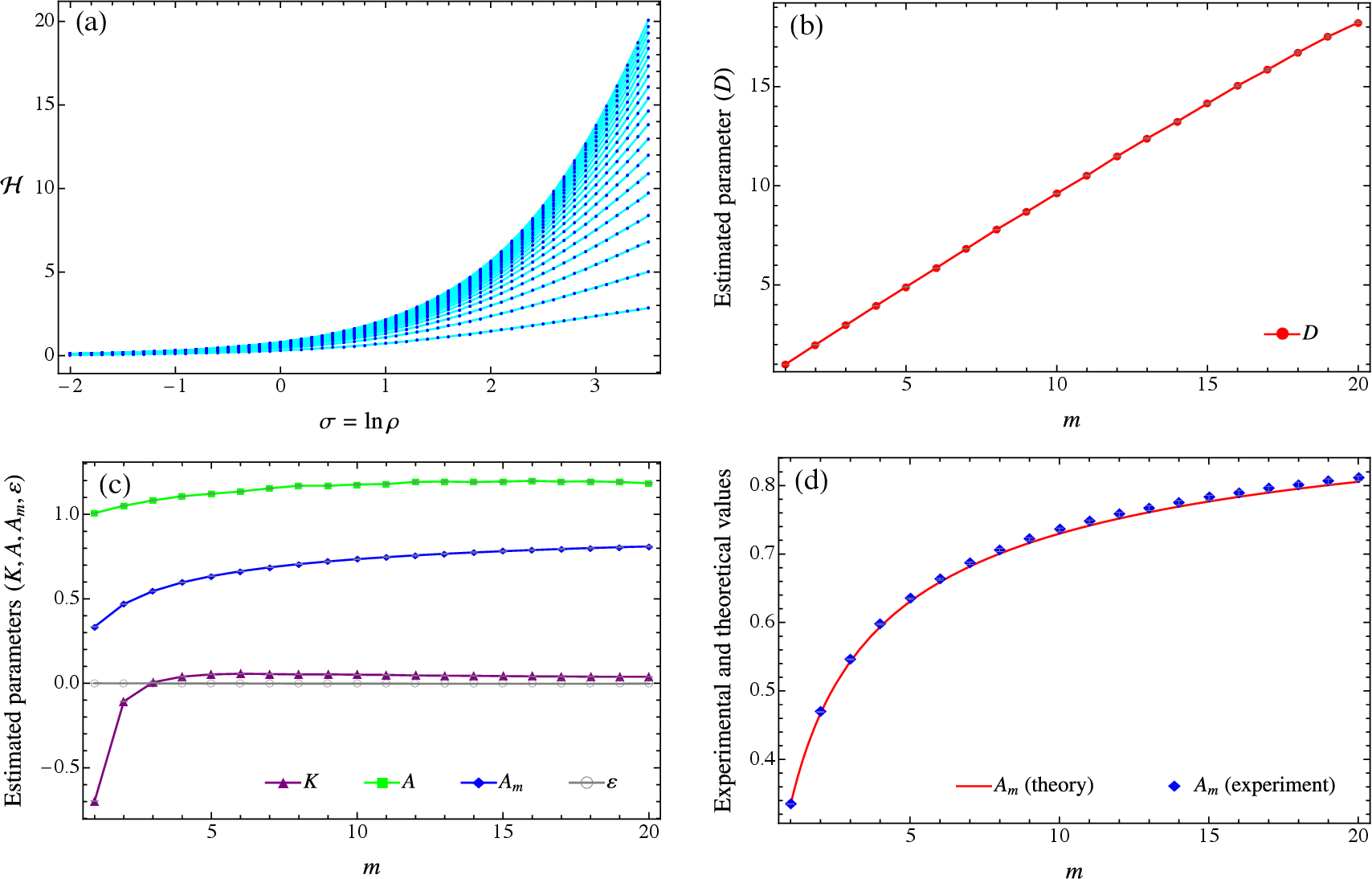}
\caption{%
The estimation results of the continuous uniform random variable case.
Panel (a) shows the graphs of the extended R\'{e}nyi entropy $\cH_{m}$ versus $\sig=\ln\rho$ for increasing values of the embedding dimension, from $m=1$ (bottom) to $m=20$ (top).
Panels (b) and (c) show the estimation results of $\cD^{*}$ and $\{\cK^{*},\cA^{*},\cA_{m}\}$, respectively, based on the hypergeometric estimator.
The error term $\varepsilon$ remains zero for all dimensions.
All the $p$-values are negligibly small.
Panel (d) shows the comparison between the theoretical predictions and the experimental results regarding the values of $\cA_{m}$ for increasing values of $m$.
In Panels (b)--(d), standard errors are indicated by capped vertical lines.
\label{fig:uniform}}
\end{figure}
\twocolumngrid

\subsubsection{Uniform random variable}

To check the validity of the method, let us first apply it to the case of continuous uniform random variables discussed earlier (Fig.~\ref{fig:topolo}(a)), with $n=10,000$.
The results are summarised in Fig.~\ref{fig:uniform}.
Panel (a) shows the graphs of the extended R\'{e}nyi entropy $\cH_{m}^{\mathrm{uni}}$ versus $\sig=\ln\rho$ for increasing values of the embedding dimension, from $m=1$ (bottom) to $m=20$ (top).
The resulting curve profiles are in contrast to those for the strange attractor cases (Fig.~\ref{fig:wing}; see also Fig.~\ref{fig:strangers} discussed later).
Panel (b) shows the estimation results of $\hat{\cD}_{\mathrm{uni}}(m)$, which evidently does not approach a constant but increases as $\sim m$.
This result agrees with the expectation from the theory; see the argument below Eq.~(\ref{M_uni}).
It is also intuitively clear as the probability density spreads out uniformly over the $m$-dimensional embedding space.
Panel (c) shows the estimation results of $\hat{\cK}_{\mathrm{uni}}(m)$, $\hat{\cA}_{\mathrm{uni}}(m)$ and $\hat{\cA}_{m}^{\mathrm{uni}}$.
Here, the notion of \q{correlation entropy} is not well-defined in the usual sense because the scaling law (\ref{scaling:sig_DK}) does not hold for the current uniform random variable.
For fixed finite $m$, the value of $\cK_{\mathrm{uni}}^{*}$ defined formally by Eq.~(\ref{K*0}) goes to infinity as $\sig\to\infty$, while for fixed finite $\sig$, it goes to zero as $m\to\infty$.\footnote{Note that $\cM_{m}^{\mathrm{uni}}(\rho)\big|_{m\gg 1}\sim m\rho^{-2m}\big(2\Gamma(2m,0,\rho)+\sqrt{\pi/m}\Gamma(2m+1,0,\rho)\big)$, where $\Gamma(a,x_{0},x_{1})\coloneqq \Gamma(a,x_{0})-\Gamma(a,x_{1})$ with $\Gamma(a,x)=\int_{z}^{\infty}t^{a-1}e^{-t}\dd t$ the upper incomplete gamma function.}
Because both the values of $\sig_{2}$ and $m$ were necessarily finite for an experimental setting, the values of $\hat{\cK}_{\mathrm{uni}}(m)$ resulted in slightly positive values.
Also, the values of $\hat{\cA}_{\mathrm{uni}}(m)$ were slightly larger than the theoretical value ($\cA^{*}_{\mathrm{uni}}=1$) to satisfy the relation (\ref{A_uni}) under Eq.~(\ref{A-Am}) for finite $m$ and positive $\hat{\cK}_{\mathrm{uni}}^{*}$.
We observed that $\hat{\cA}_{\mathrm{uni}}^{*}$ approaches close to the theoretical value by narrowing the fitting window $[\sig_{0},\sig_{2}]$, i.e., with a smaller value of the right endpoint $\sig_{2}$.
The error term $\varepsilon$ arising from the regression remained zero for all dimensions, and all the $p$-values were negligibly small.
Panel (d) shows the comparison between the theoretical prediction and the experiment results regarding the probabilistic mean distance $\cA_{m}^{\mathrm{uni}}$.
The estimated value at each $m$ agreed with the theoretical value given by Eq.~(\ref{Am_uni}).
Again, we observed that the experimental results match more accurately with the theoretical values with a narrower fitting window.

Note that the simulation results for the discretised straight-line case discussed earlier (Fig.~\ref{fig:topolo}(b)) were also obtained in the $m=1$ case of the current uniform random distribution case.
We obtained $\hat{\cA}_{1}^{\mathrm{uni}}=0.3349\pm 0.0002$, which agreed with the theoretical value of $\cA_{\mathrm{line}}^{*}=1/3$ (see Eq.~(\ref{A_1=1/3})).
Also, the result that $\hat{\cK}_{\mathrm{uni}}(m=1)<0$ indicated that this system was perfectly predictable without any diverging behaviours.

\subsubsection{Discrete maps}

\begin{figure}[tp]
\centering
\vspace{-2em}
\includegraphics[width=0.87\textwidth]{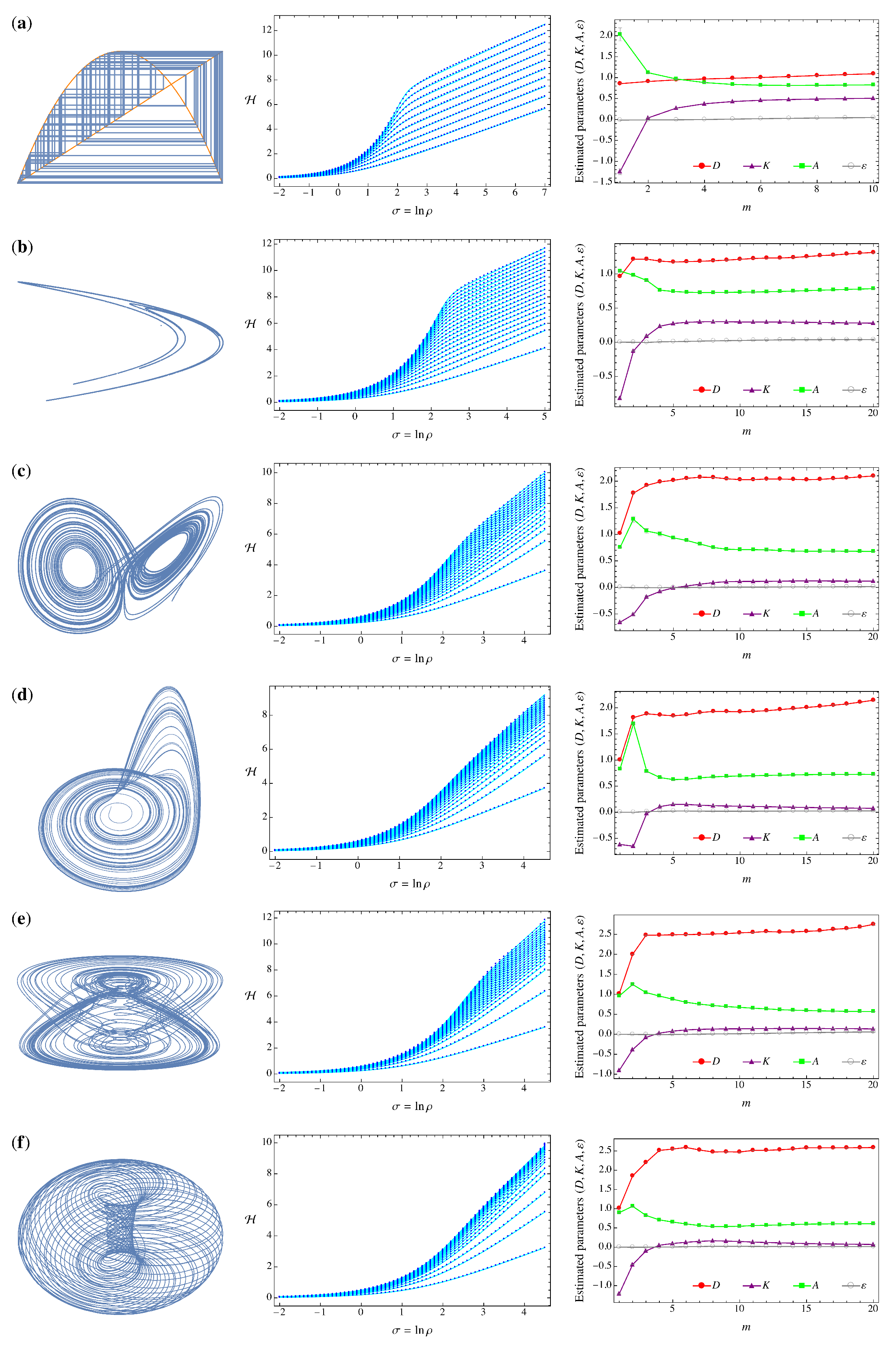}
\vspace{-1em}
\caption{%
Estimation results of six strange attractors; see Table \ref{tab:summary} for the summary statistics.
The left column shows the attractors' appearances.
The middle column shows the graphs of the extended R\'{e}nyi entropy $\cH_{m}$ versus $\sig=\ln\rho$ for increasing values of the embedding dimension, from $m=1$ (bottom) to $m=20$ (top).
The right column shows the estimation results of the three chaotic invariants.
The error term $\varepsilon$ remains zero for all dimensions.
Standard errors are indicated by capped vertical lines.
\label{fig:strangers}}
\end{figure}

Now, we present the results for strange attractors.
Table \ref{tab:summary} summarises the simulation setups and the estimation results.
We begin with the discrete systems.
First, logistic map (Fig.~\ref{fig:strangers}(a)) is defined by 
\begin{equation}\label{eq:Logistic}
x_{i+1}=Ax_{i}(1-x_{i})\,.
\end{equation}
We used parameters of $A=4$, with initial condition $x_{1}=0.1$ and the averaging range of embedding dimensions $m=4$--$8$.
We obtained the estimates of $\hat{\cD}^{*}=1.01\pm 0.02$ for the correlation dimension, close to the theoretical value $D_{2}=D_{\mathrm{KY}}=1$ (exact value), also in agreement with previously reported results, e.g., $D_{2}= 1.016\pm 0.023$ \cite{Sprott01}.
We also obtained $\hat{\cK}^{*}=0.45\pm 0.02$ and $\hat{\cA}^{*}=0.83\pm 0.01$ for the correlation entropy and concentration, respectively.
To further check the validity of the estimator, we compared the estimation result of $\hat{\cA}_{1}=0.406$ at $m=1$ with the analytical result.
This consistency check was possible because the physical invariant measure for the $A=4$ logistic map is known analytically, which is given by $\dd\mu(x)=dx\big/\pi\sqrt{x(1-x)}$.
The probabilistic mean distance evaluated in the one-dimensional embedding space is given by $\iint\abs{x-y}\dd\mu(x)\dd\mu(y)=4/\pi^{2}\approx 0.405$, closely reproducing the above experimental result.

The second example is H\'{e}non map \cite{Henon76} (Fig.~\ref{fig:strangers}(b)) defined by 
\begin{equation}\label{eq:Henon}
x_{i+1}=1-ax_{i}^{2}+y_{i}\,,~\, 
y_{i+1}=bx_{i}\,.
\end{equation}
We used the standard parameters of $a=1.4$ and $b=0.3$, with initial conditions $(x_{1},y_{1})=(0.5,0.1)$ and the averaging dimension range $m=5$--$20$.
We obtained the estimates of $\hat{\cD}^{*}=1.24\pm 0.01$ (close to $D_{\mathrm{KY}}=1.258$) and $\hat{\cK}^{*}=0.291\pm 0.002$, both in agreement with previously reported results, e.g., $D_{2}=1.25\pm 0.02$ \cite{Grassberger83b}, $1.227\pm 0.011$ \cite{Yu00}, $1.220\pm 0.036$ \cite{Sprott01} and $1.23\pm 0.1$ \cite{Harikrishnan06} for the correlation dimension, and $K_{2}= 0.29\pm 0.1$ \cite{Frank93} and $0.301\pm 0.003$ \cite{Yu00} for the correlation entropy.
We also obtained $\hat{\cA}^{*}=0.749\pm 0.005$.

In the middle column of Fig.~\ref{fig:strangers}(b), the curve of $\cH_{m=1}$ (bottom) evidently deviates from the parallel linear segments at $\rho\gg 1$.
Specifically, we obtained $\hat{\cD}(m=1)=0.9642\pm 0.0004$, which was significantly lower than the other values of $\hat{\cD}(m\geq 2)$, or the resulting $\hat{\cD}^{*}$.
This deviation occurred because the embedding dimension ($m=1$) was not large enough to unfold the attractor's structure fully.
More generally, the curve of $\hat{\cD}(m)$ against $m$ could become flat around $\cD^{*}$ only for $m>\cD^{*}$.
Further, to check the invariance of these three characteristics under the change of coordinates, we also conducted the estimation by using the $y$-projected time series ($\{y_{i}\}_{i=1,\,\dots,\,n}$).
The results were precisely the same as those obtained in the $x$-projected case given above, ensuring the invariance property of $\{\cD^{*},\,\cK^{*},\,\cA^{*}\}$.\\

\onecolumngrid

\twocolumngrid

\subsubsection{Continuous flows}

Next, we move on to the continuous systems. 
First, Lorenz attractor \cite{Lorenz63} (Fig.~\ref{fig:strangers}(c)) is defined by
\begin{equation}\label{eq:Lorenz}
\dot{x}=\bar{\sigma} (y-x)\,,~\, 
\dot{y}=x(\bar{\rho}-z)-y\,,~\, 
\dot{z}=xy-{\beta} z\,,
\end{equation}
where a dot denotes derivative with respect to time, e.g., $\dot{x}\equiv\partial_{t}x$.
We used the standard parameters of $\bar{\sigma}=10$, $\bar{\rho}=28$ and ${\beta}=8/3$, with initial conditions $(x(0),y(0),z(0))=(10,10,10)$ and trajectory length $n=245$.
We obtained the estimates of $\hat{\cD}^{*}=2.048\pm 0.007$ (close to $D_{\mathrm{KY}}=2.062$), in agreement with previously reported results; e.g., $D_{2}=2.049\pm 0.096$ \cite{Sprott01} and $2.03\pm 0.16$ \cite{Harikrishnan06}, based on the same parameter set as ours.
We also obtained $\hat{\cK}^{*}=0.118\pm 0.001$ and $\hat{\cA}^{*}=0.690\pm 0.004$.

R\"{o}ssler attractor \cite{Rossler76} (Fig.~\ref{fig:strangers}(d)) is defined by
\begin{equation}\label{eq:Rossler}
\dot{x}=-y-z\,,~\, 
\dot{y}=x+ay\,,~\, 
\dot{z}=b+z(x-c)\,.
\end{equation}
We used parameters of $a=b=0.2$ and $c=5.7$ with initial conditions $(x(0),y(0),z(0))=(1,1,1)$ and trajectory length $n=341$.
We obtained the estimates of $\hat{\cD}^{*}=2.02\pm 0.02$ (close to $D_{\mathrm{KY}}=2.013$) and $\hat{\cK}^{*}=0.093\pm 0.004$, in agreement with previously reported results; e.g., $D_{2}=1.986\pm 0.078$ \cite{Sprott01} and $2.06\pm 0.06$ \cite{Romano04} for the correlation dimension, and $K_{2}=0.067\pm 0.007$ \cite{Romano04} for the correlation entropy, based on the same parameter set as ours.
In addition, we obtained $\hat{\cA}^{*}=0.716\pm 0.004$.

Duffing--Ueda attractor (also commonly known as Japanese attractor) \cite{Ueda79} (Fig.~\ref{fig:strangers}(e)) is defined by
\begin{equation}\label{eq:Ueda}
\ddot{x}+k\dot{x}+x^{3}=B\cos t\,.
\end{equation}
We used parameters of $k=0.05$ and $B=7.5$, with initial conditions $x(0)=2.5$ and trajectory length $n=375$.
We obtained the estimates of $\hat{\cD}^{*}=2.60\pm 0.02$ (close to $D_{\mathrm{KY}}=2.674$), in agreement with previously reported results, e.g., $D_{2}=2.675\pm 0.132$ \cite{Sprott01} and $2.59\pm 0.1$ \cite{Harikrishnan06}, based on the same parameter set as ours.
We also obtained $\hat{\cK}^{*}=0.141\pm 0.001$ and $\hat{\cA}^{*}=0.60\pm 0.01$.

Finally, Langford attractor \cite{Langford84} (also commonly known as Aizawa attractor) (Fig.~\ref{fig:strangers}(f)) is defined by
\begin{align}\label{eq:Langford}
&\dot{x}=(z-\beta)x-\omega y\,,~\, 
\dot{y}=\omega x+(z-\beta)y\,,~\, \no\\
&\dot{z}=\lam+\alpha z-\f{z^{3}}{3}+b\big(x^{2}+y^{2}\big)(1+\bar{\rho} z)+\varepsilon z x^{3}\,.
\end{align}
We used parameters of $\al=0.95$, $\be=0.7$, $\omega=3.5$, $\lam=0.6$, $\bar{\rho}=0.25$ and $\varepsilon=0.01$, with initial conditions $(x(0),y(0),z(0))=(0.1,1,0)$ and trajectory length $n=330$.
We obtained the estimates of $\hat{\cD}^{*}=2.55\pm 0.01$, $\hat{\cK}^{*}=0.101\pm 0.008$ and $\hat{\cA}^{*}=0.589\pm 0.007$.
Although we have been unable to find relevant previous studies investigating the invariants of this system, it is interesting to see if the above results are consistent with theoretical values or the estimation results derived from the GP-based methods or other sophisticated techniques.

\onecolumngrid
\begin{table*}[t]
\centering
\caption{Summary of the estimates of $\cD^{*}$, $\cK^{*}$ and $\cA^{*}$ based on the hypergeometric estimator for six strange attractor systems.
All the $p$-values are negligibly small.
\label{tab:summary}} 
{\small
\begin{xtabular*}{0.95\textwidth}{cl@{\hspace{2em}}l@{\hspace{2em}}l@{\hspace{2em}}l@{\hspace{2em}}l@{\hspace{2em}}l@{\hspace{2em}}l@{\hspace{0em}}}
\toprule\\[-1em]
{} & Strange attractor system & $n$ & $\sigma_{2}$ & $m$ & $\hat{\cD}^{*}$ & $\hat{\cK}^{*}$ & $\hat{\cA}^{*}$ \\
\hline\\[-3mm]
a. & Logistic map, Eq.~(\ref{eq:Logistic}) & 10,000 & 7.0 & [4,8] & $1.01\pm 0.02$ & $0.45\pm 0.02$ & $0.83\pm 0.01$ \\
b. & H\'{e}non map, Eqs.~(\ref{eq:Henon}) & 10,000 & 5.0 & [5,20] & $1.24\pm 0.01$ & $0.291\pm 0.002$ & $0.749\pm 0.005$ \\
c. & Lorenz attractor, Eqs.~(\ref{eq:Lorenz}) & \multicolumn{1}{c}{245} & 4.5 & [10,20] & $2.048\pm 0.007$ & $0.118\pm 0.001$ & $0.690\pm 0.004$ \\
d. & R\"{o}ssler attractor, Eqs.~(\ref{eq:Rossler}) & \multicolumn{1}{c}{341} & 4.5 & [10,20] & $2.02\pm 0.02$ & $0.093\pm 0.004$ & $0.716\pm 0.004$ \\
e. & Duffing--Ueda attractor, Eq.~(\ref{eq:Ueda}) & \multicolumn{1}{c}{375} & 4.5 & [10,20] & $2.60\pm 0.02$ & $0.141\pm 0.001$ & $0.60\pm 0.01$ \\
f. & Langford attractor, Eq.~(\ref{eq:Langford}) & \multicolumn{1}{c}{330} & 4.5 & [10,20] & $2.55\pm 0.01$ & $0.101\pm 0.008$ & $0.589\pm 0.007$ \\
\hline
\end{xtabular*}
}
\end{table*}
\twocolumngrid

\section{Discussion: Hypergeometric structure behind chaos\label{sec:discussion}}

The simulation results presented in the previous section provide evidence that the extended correlation integral (\ref{M_m}) can be closely approximated by Kummer's confluent hypergeometric function of the form (\ref{M:est}).
However, it remains unclear how one can justify the use of Kummer's function from a theoretical viewpoint beyond intuition. 
In this section, we explain the rationale of Kummer's function being natural, albeit without mathematically rigorous proof. 
In so doing, we present a conjecture on the PDF of the distance between the reconstruction vectors ($r_{ij}= \|\bX_{i}-\bX_{j}\|$), evaluated by the underlying probability measure.

A key fact here is that Kummer's function is the moment-generating function of a beta distribution.
This fact and the above-observed accurate simulation results suggest that the PDF of $r_{ij}$ is closely related to beta distribution functions.
We conjecture that the PDF of $r_{ij}$ in the $m$-dimensional embedding space can be described by a superposition of incomplete beta distributions,
\begin{equation}
f_{m}(r)\approx\int_{s\in\mathscr{S}}w(s)g_{m}^{(s)}(r)\dd s\,.
\end{equation}
Here, $\mathscr{S}$ denotes the set of labels attached to each incomplete beta distribution, $w(s)\in [0,1]$ is the weight function satisfying $\int_{s\in\mathscr{S}}w(s)\dd s=1$, and 
\begin{align}
\resizebox{.48\textwidth}{!} 
{$
g_{m}^{(s)}(r)\coloneqq
\left\{
\begin{aligned}
&\, \f{\big(\f{r}{\xi(s)}\big)^{\al(s)-1}\big(1-\f{r}{\xi(s)}\big)^{\be_{m}(s)-1}}{\mathrm{B}(\al(s),\be_{m}(s))} && (0\leq r<\xi(s))\,,\\ 
&\,\quad 0 && (\xi(s)<r\leq 1)\,,
\end{aligned}
\right.
$}
\end{align}
with parameters $\al(s)>0$, $\be_{m}(s)>0$ and $\xi(s)\in (0,1]$.
If $\xi=1$, this PDF reduces to that of the standard beta distribution with shape parameters $\al(s)$ and $\be_{m}(s)$.
If $\mathscr{S}$ is a discrete set composed of elements $\{s_{k}\}_{k=1,\,2,\,\dots}$, the weight function is given by a weighted sum of Dirac delta functions, $w(s)=\sum_{k}w_{k}\delta(s-s_{k})$ with $w_{k}\in [0,1]$, $\sum_{k}w_{k}=1$.
Then, the extended correlation integral (\ref{M_m}) is computed as
\begin{align}\label{conj:M}
\cM_{m}(\rho)&=\int_{0}^{1}f_{m}(r)\,e^{-\rho r}\dd r\no\\
&=\int_{s\in\mathscr{S}}w(s)\M[\al(s),\al(s)+\be_{m}(s);-\xi(s)\rho]\dd s\,.
\end{align}
The probabilistic mean distance is given by the first moment of $\cM_{m}(\rho)$ about $\rho=0$, i.e.,
\begin{equation}
\cA_{m}
=-\lim_{\rho\to 0}\f{d\cM_{m}(\rho)}{d\rho}
=\int_{s\in\mathscr{S}}\f{w(s)\xi(s)\al(s)}{\al(s)+\be_{m}(s)}\dd s\,,
\end{equation}
satisfying the small-$\rho$ scaling law of Eq.~(\ref{scaling:A}).
In the large-$\rho$ limit, it can be shown that
\begin{align}\label{sc:inf:M}
\cM_{m}(\rho)\, &\sim\, \int_{s\in\mathscr{S}}w(s)\f{\Gamma\big(\al(s)+\be_{m}(s)\big)}{\Gamma\big(\be_{m}(s)\big)}\rho^{-\al(s)}\dd s\,,
\end{align}
where we have again used the transformation formula and the large-$x$ expansion formula for Kummer's confluent hypergeometric function.
Note that the right-hand side of Eq.~(\ref{sc:inf:M}) is dominated by the term proportional to $\rho^{-\al(s^{*})}$ with $s^{*}\coloneqq\argmin_{s}\al(s)$.
Hence, in view of the scaling law (\ref{scaling:DK}), we may identify 
\begin{equation}
\al(s^{*})\equiv\cD^{*}\,,\quad \be_{m}(s^{*})\equiv\cD^{*}e^{-\cK^{*}m\tau}\,.
\end{equation}
With all these observations, we conjecture that the geometrical and dynamical information of the probability measure $\mu_{m}$ of a strange attractor can be extracted from the spectrum of $\{\al(s),\,\be_{m}(s),\,\xi(s),\,w(s)\}_{s\in\mathscr{S}}$ associated with the hypergeometric description of the correlation integral (\ref{conj:M}).
We also conjecture that this feature reflects the multifractal property of the chaotic system, each monofractal being labeled by $s\in\mathscr{S}$.

\section{Summary and conclusions\label{sec:summary}}

A new description of strange attractor systems through three chaotic invariants was provided.
They were the well-known correlation dimension ($\cD^{*}$) and the correlation entropy $(\cK^{*})$, both having attracted attention over the past decades, and a new characteristic called the correlation concentration ($\cA^{*}$) introduced in the present study.
The correlation concentration was defined as the normalised mean distance between the reconstruction vectors $\{\bX_{i}\}$, evaluated by the probability measure ($\mu_{m}$) on the infinite-dimensional ($m\to\infty$) embedding space.
It was shown that these three invariants determine the scaling behaviour of the system's R\'{e}nyi-type extended entropy ($\cH_m$) defined on the $m$-dimensional embedding space, whose theoretical foundation has been provided in Ref.~\cite{Okamura20}.

The extended entropy was derived from the modified correlation integral (\ref{M_m}), in which the Heaviside function used in the GP algorithm \cite{Grassberger83a,Grassberger83b} was replaced by an exponential function of the form $e^{-\rho\|\bX_{i}-\bX_{j}\|}$ with $\rho>0$ the gauge parameter.
Explicitly, it was formulated as $\cH_{m}(\rho)=-\ln\iint\dd\mu_{m}(\bX_{i},\bX_{j})\,e^{-\rho\|\bX_{i}-\bX_{j}\|}$.
Due to this kernel structure, the extended entropy was capable of smoothly interpolating between the \q{microscopic} scale ($\rho\to\infty$) and the \q{macroscopic} scale ($\rho\to 0$) of the attractor.
The plot of $\cH_{m}$ versus $\sig=\ln\rho$ exhibited a wing-shaped profile as shown in Fig.~\ref{fig:wing}.
At $\rho\gg 1$, the slope of the entropy curve represented $\cD^{*}$, while the vertical gap between the consecutive linear segments represented $\cK^{*}$.
In addition, at $\rho\ll 1$, the exponential curvature was controlled by $\cA^{*}$ and $\cK^{*}$.

The extended entropy function was modelled by Kummer's confluent hypergeometric function, whose parameters involved the three chaotic invariants $\{\cD^{*},\,\cK^{*},\,\cA^{*}\}$.
Explicitly, the model function was given by $\widetilde\cH_{m}(\sig)=-\ln\M[\cD^{*},\cD^{*}(1+e^{-\cK^{*} m\tau});-\cA^{*} e^{\sig}]$ with $\tau$ the sampling time of the time series.
It reproduced the required scaling behaviours in both limits ($\sig\to\pm\infty$) of the scale parameter, hence was suitable for fitting the observed time series data throughout the parameter region.

The proposed method was verified using experimental chaotic time series generated by six strange attractor systems.
They were two discrete maps (logistic and H\'{e}non maps) and four continuous flows (Lorenz, R\"{o}ssler, Duffing--Ueda and Langford attractors).
The hypergeometric estimator $\widetilde\cH_{m}$ was used to estimate the three chaotic invariants simultaneously via nonlinear regression analysis, without needing separate estimations for each invariant.
Except for the Langford attractor, for which relevant previous studies were unavailable, our simulation results closely reproduced the known theoretical or/and experimental values of the correlation dimension and entropy, while also providing new knowledge on the estimates of the correlation concentration.

Further, based on the experimental evidence and analytical observations, a conjecture was made concerning the relation between the chaotic invariants and the underlying probability measure.
The PDF of the distances between the reconstruction vectors was conjectured to be described by the PDF of mixed incomplete beta distributions whose parameters involved the spectrum of the chaotic invariants.
All these results and conclusions provided an improved and deeper understanding of the geometrical and dynamical properties of chaotic systems.
Its connection to the extreme value theory (e.g., \cite{Caby19}) would also be an interesting subject to explore.

Still, it should be emphasised that all the above concepts have been formulated in a purely deterministic setting, driven by theoretical motivations and considerations.
Therefore, the findings and implications of this paper would be only valid for infinite and noise-free time series of a dynamical system, at least for now.
In the real world, noise is inherently present in almost all observational time series, and only short time series are available in general.
Consequently, the finite resolution and duration of the time series will break the invariance property of the correlation dimension, entropy and concentration.
Future works may follow to address these limitations by extending the currently developed framework to be applied to the real-world time series arising from various natural phenomena.

\begin{acknowledgments}
The author would like to thank Ryo Suzuki, Shuhei Mano and the two anonymous reviewers of \textit{Chaos, Solitons \& Fractals} for their valuable comments.
This research did not receive any specific grant from funding agencies in the public, commercial, or not-for-profit sectors.
The views and conclusions contained herein are those of the author and should not be interpreted as necessarily representing the official policies or endorsements, either expressed or implied, of the organisation to which the author is affiliated.
\end{acknowledgments}

\section*{Data Availability}
The datasets and figures generated and/or analysed during this study can be found in the Zenodo repository at \url{https://doi.org/10.5281/ZENODO.7791696}.

\appendix

\renewcommand{\thefigure}{\Alph{section}\arabic{figure}}
\setcounter{section}{0}
\setcounter{figure}{0}
\setcounter{table}{0}
\setcounter{equation}{0}

\vspace{1.0cm}
\section{Correlation concentration in simple systems\label{app:cc}}

\begin{figure}[t]
 \centering
\includegraphics[width=0.43\textwidth]{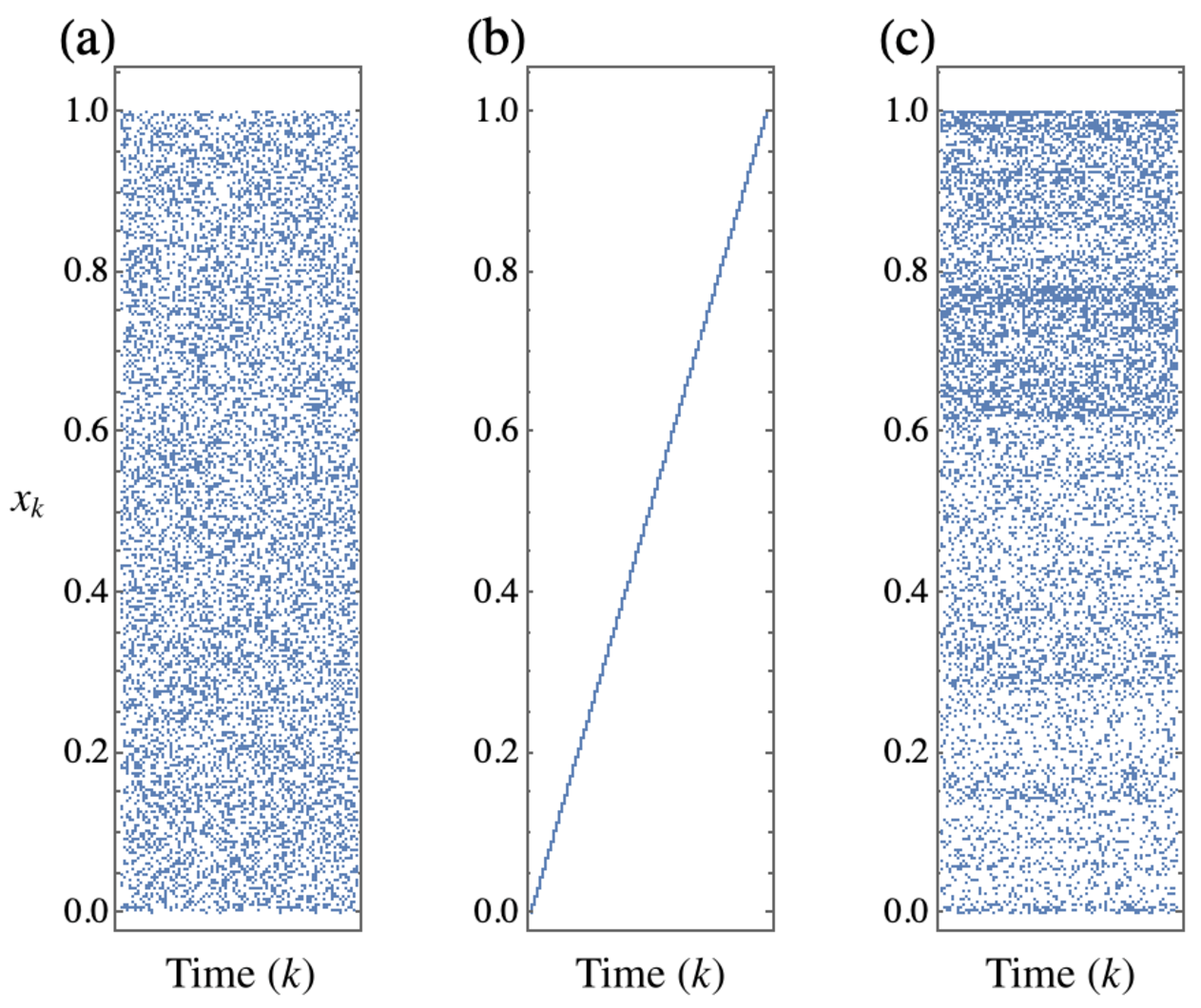}
\caption{%
Scatterplots of $\{x_{k}\}_{k=1,\,\dots,\,n}$ against $k$ (time variable) with $n=10,000$ for three different topological situations.
(a) Discrete uniform random variable; (b) Discretised straight line defined by $x_{k}=k/n$; (c) The (rescaled) $x$-component of H\'{e}non map defined by $(x_{k},y_{k})\mapsto (x_{k+1},y_{k+1})=(1-1.4x_{k}^{2}+y_{k},0.3x_{k})$ with initial conditions $(x_{1},y_{1})=(0.5,0.1)$.
\label{fig:topolo}}
\end{figure}

It is instructive to see by simple examples how the correlation concentration is computed for given $m$ and behaves as $m$ increases to infinity, depending on the topological nature of dynamical systems.
Let us first consider the case of continuous random variables $\{x_{i}\}$ and $\{x_{j}\}$, which are independently and uniformly distributed on the interval $[0,1]$; see Fig.~\ref{fig:topolo}(a) for a discretised illustration.
A short calculation shows that the probability distribution function (PDF) of $r_{ij}\coloneqq\|\bX_{i}-\bX_{j}\|$ in a $m$-dimensional embedding space is given by 
\begin{equation}\label{PDF_uni}
f_{m}^{\mathrm{uni}}(r)=2m(1-r)r^{m-1}(2-r)^{m-1}\,.
\end{equation}
This result can be obtained by noticing that $P(\|\bX_{i}-\bX_{j}\|<r)=\prod_{k=0}^{m-1}P(\abs{x_{i+k}-x_{j+k}}<r)=\big(1-(1-r)^{2}\big)^{m}$ and taking its derivative with respect to $r$.
The expectation value of $r_{ij}$ at finite $m$ is then given by
\begin{equation}\label{Am_uni}
\cA_{m}^{\mathrm{uni}}=\int_{0}^{1}r f_{m}^{\mathrm{uni}}(r)\dd r
=1-\f{m\mathrm{B}\big(\f{1}{2},m\big)}{2m+1}\,,
\end{equation}
where $\mathrm{B}(\al,\be)=\int_{0}^{1}t^{\alpha-1}(1-t)^{\beta-1}\dd t$ is the Euler beta function.
The large-$m$ expansion of Eq.~(\ref{Am_uni}) is given by $\cA_{m}^{\mathrm{uni}}\sim 1-\sqrt{\pi/4m}$, and therefore, the correlation concentration of the continuous uniform distribution is given by
\begin{equation}\label{A_uni}
\cA^{*}_{\mathrm{uni}}=\lim_{m\to\infty}\cA_{m}^{\mathrm{uni}}=1\,.
\end{equation}
This gives the upper bound of the correlation concentration for any system.
More generally, the $k$-th moment of the PDF (\ref{PDF_uni}) can be calculated as
\begin{align}
a_{k,m}^{\mathrm{uni}}&=\int_{0}^{1}r^{k} f_{m}^{\mathrm{uni}}(r)\dd r\no\\
&=\f{2m\big(1-2^{2m+k-1}s\mathrm{B}_{\f{1}{2}}(m+k,m)\big)}{2m+k}\,,
\end{align}
where $\mathrm{B}_{x}(\al,\be)=\int_{0}^{x}t^{\alpha-1}(1-t)^{\beta-1}\dd t$ is the incomplete beta function.
By definition, $a_{1,m}^{\mathrm{uni}}\equiv\cA_{m}^{\mathrm{uni}}$.
Subsequently, the extended correlation integral for the continuous uniform variables is given by
\begin{equation}\label{M_uni}
\cM_{m}^{\mathrm{uni}}(\rho)=\sum_{k=0}^{\infty}a_{k,m}^{\mathrm{uni}}\f{(-\rho)^{k}}{k!}\,.
\end{equation}
It can be shown that this function behaves as $\cM_{m}^{\mathrm{uni}}(\rho)\sim\rho^{-m}$ as $\rho\to\infty$, that is, the exponent on the gauge parameter scales as the embedding dimension.
This is clearly a distinct asymptotic behaviour from the strange attractor case displayed in Eq.~(\ref{scaling:DK}), in which the magnitude of the exponent on the gauge parameter is constant---the correlation dimension.

As the next example, let us consider the simplest deterministic case, i.e., a straight-line trajectory.
See Fig.~\ref{fig:topolo}(b) for a discretised illustration.
In this case, the extended correlation integral is calculated as
\begin{equation}
\cM_{m}^{\mathrm{line}}(\rho)=\int_{0}^{1}\!\!\int_{0}^{1} e^{-\rho\abs{x-y}}\dd x\dd y
=\f{2\ko{e^{-\rho}+\rho-1}}{\rho^{2}}\,,
\end{equation}
which is of course independent of $m$.
The correlation concentration is computed as
\begin{equation}\label{A_1=1/3}
\cA^{*}_{\mathrm{line}}=\cA_{m}^{\mathrm{line}}=-\lim_{\rho\to 0}\f{d\cM_{m}^{\mathrm{line}}(\rho)}{d\rho}=\f{1}{3}\,.
\end{equation}
By construction, these results agree with the $m=1$ case of the uniform random variable case, i.e., $\cM_{m}^{\mathrm{line}}=\cM_{1}^{\mathrm{uni}}$ and $\cA_{m}^{\mathrm{line}}=\cA_{1}^{\mathrm{uni}}$.

\section{Derivation of Eq.~(\ref{lim:H_DK})\label{app:derivation}}

Here, we show how Eq.~(\ref{lim:H_DK}) in the main text is obtained.
Using the transformation formula $\M(a,b;x)=e^{x}\M(b-a,b;-x)$ (see 13.2.39 of \cite{Mathews22}), the Kummer's function part of Eq.~(\ref{M:est}) translates to 
\begin{align}\label{app:Eq1}
&\M[\cD^{*},\cD^{*}(1+e^{-\cK^{*} m\tau});-\cA^{*} e^{\sig}]\no\\
&~{}=e^{-\cA^{*} e^{\sig}}\M[\cD^{*}e^{-\cK^{*} m\tau},\cD^{*}(1+e^{-\cK^{*} m\tau});\cA^{*} e^{\sig}]\,.
\end{align}
In the $\sigma\to\infty$ limit, the Kummer's function part of the RHS of Eq.~(\ref{app:Eq1}) behaves as
\begin{align*}
&\M[\cD^{*}e^{-\cK^{*} m\tau},\cD^{*}(1+e^{-\cK^{*} m\tau});\cA^{*} e^{\sig}]\no\\
&\quad {}\sim\, 
\f{\Gamma\big(\cD^{*}\big(1+e^{-\cK^{*}m\tau}\big)\big)}{\Gamma\big(\cD^{*}e^{-\cK^{*}m\tau}\big)}\cdot e^{\cA^{*} e^{\sig}}\cdot\ko{\cA^{*} e^{\sig}}^{-\cD^{*}}
\end{align*}
in view of the large-$x$ expansion formula (see 13.2.4 and 13.2.23 of \cite{Mathews22}), $\M(a,b;x)\sim \big(\Gamma(b)/\Gamma(a)\big)e^{x}x^{a-b}[1+\cO(x^{-1})]$.
Plugging this asymptotic form into Eq.~(\ref{app:Eq1}), we obtain the $\sigma\to\infty$ limit of Eq.~(\ref{M:est}) as
\begin{equation}\label{app:Eq2}
\widetilde{\cH}_{m}(\sig)\stackrel{\sig\to\infty}{\sim}\, \cD^{*}\sig+\ln\kko{\f{\cA^{*}{}^{\cD^{*}}\Gamma\big(\cD^{*}e^{-\cK^{*}m\tau}\big)}{\Gamma\big(\cD^{*}\big(1+e^{-\cK^{*}m\tau}\big)\big)}}\,.
\end{equation}
Finally, using the asymptotic behaviour of the gamma function, $\Gamma(x)=x^{-1}+\cO(1)$ as $x\to 0$, the above expression further reduces to
\begin{equation}\label{app:Eq3}
\widetilde{\cH}_{m}(\sig)\,\sim\,\cD^{*}\sig+\cK^{*}m\tau +\text{const.}
\quad\text{as}~\, \sig\to \infty,\, m\to \infty\,,
\end{equation}
which was to be shown.

\bibliographystyle{apsrev4-2}

\providecommand{\noopsort}[1]{}\providecommand{\singleletter}[1]{#1}%

\end{document}